\documentclass[journal]{IEEEtran}
\ifCLASSINFOpdf
\else
\fi

\usepackage{amsmath}
\usepackage[ruled]{algorithm2e}
\usepackage{array}
\usepackage{url}
\usepackage{graphicx}
\usepackage{multicol}
\usepackage{multirow}
\usepackage{xcolor}
\usepackage{hyperref}
\usepackage{pifont}
\begin{document}

% \begin{frontmatter}

\title{Cross-subject Brain Functional Connectivity Analysis for Multi-task Cognitive State Evaluation}

\author{Jun~Chen,~\IEEEmembership{Member,~IEEE,}
        Anqi~Chen,
        Bingkun~Jiang,
        Mohammad~S.~Obaidat,~\IEEEmembership{Life~Fellow,~IEEE,} \\
        Ni~Li,
        Xinyu~Zhang*,~\IEEEmembership{Member,~IEEE}
        % Vincent~CS~Lee,~\IEEEmembership{Senior Member,~IEEE,}

% \author{Jun~Chen,
%         Anqi~Chen,
%         Bingkun~Jiang,
%         Mohammad~S.~Obaidat,
%         Ni~Li,
%         Xinyu~Zhang*
        
\thanks{*Correspondence: X. Zhang is with the School of Electronics and Information, Northwestern Polytechnical University, Xi'an, Shaanxi, $710072$ China, Email: xinyu.zhang@nwpu.edu.cn.}

\thanks{J. Chen is an Associate Professor with the School of Electronics and Information, Northwestern Polytechnical University, Xi'an $710072$, China; Chongqing Institute for Brain and Intelligence, Guangyang Bay Laboratory, Chongqing $400064$, China. Email: junchen@nwpu.edu.cn}

\thanks{A. Chen and B. Jiang are with the School of Electronics and Information, Northwestern Polytechnical University, Xi'an $710072$, China.
Email: chenanqi@mail.nwpu.edu.cn, kun@nwpu.edu.cn}

\thanks{Mohammad S. Obaidat is a Distinguished Professor at the King Abdullah II School of Information Technology, The University of Jordan, Amman $11942$, Jordan and School of Computer and Communication Engineering, University of Science and Technology Beijing, Beijing $100083$, China, Department of Computational Intelligence, School of Computing, SRM University, SRM Nagar, Kattankulathur $603203$, TN, India and School of Engineering, The Amity University, Noida, UP $201301$, India. Email: msobaidat@gmail.com or m.s.obaidat@ieee.org}

\thanks{N. Li is a Professor with the School of Aeronautics, Northwestern Polytechnical University, Xi'an $710072$, China. Email: lini@nwpu.edu.cn}
}

% The paper headers
\markboth{Journal of \LaTeX\ Class Files,~Vol.~14, No.~8, July~2024}%
{Shell \MakeLowercase{\textit{et al.}}: Bare Demo of IEEEtran.cls for IEEE Journals}

% make the title area
\maketitle

\begin{abstract}
Cognition refers to the function of information perception and processing, which is the fundamental psychological essence of human beings.
It is responsible for reasoning and decision-making, while its evaluation is significant for the aviation domain in mitigating potential safety risks.
Existing studies tend to use varied methods for cognitive state evaluation yet have limitations in timeliness, generalisation, and interpretability.
Accordingly, this study adopts brain functional connectivity with electroencephalography signals to capture associations in brain regions across multiple subjects for evaluating real-time cognitive states.
Specifically, a virtual reality-based flight platform is constructed with multi-screen embedded.
Three distinctive cognitive tasks are designed and each has three degrees of difficulty.
Thirty subjects are acquired for analysis and evaluation.
The results are interpreted through different perspectives, including inner-subject and cross-subject for task-wise and gender-wise underlying brain functional connectivity.
Additionally, this study incorporates questionnaire-based, task performance-based, and physiological measure-based approaches to fairly label the trials.
A multi-class cognitive state evaluation is further conducted with the active brain connections.
Benchmarking results demonstrate that the identified brain regions have considerable influences in cognition, with a multi-class accuracy rate of $95.83\%$ surpassing existing studies.
The derived findings bring significance to understanding the dynamic relationships among human brain functional regions, cross-subject cognitive behaviours, and decision-making, which have promising practical application values.
\end{abstract}


\IEEEpeerreviewmaketitle

\section{Introduction}
\label{sec1}
\IEEEPARstart{H}{uman} brain is a sophisticated organ that generates signals among regions through the cortical area interaction, which is established through a densely functional connectivity network \cite{Mict2022}.
Such a network helps provide an in-depth understanding of driving factors affecting behaviour and cognition, meanwhile allowing flexible and adaptive allocation of resources to accomplish task demands in high expectations \cite{Cai2024}.
There has been a concurrent rise in evincing that distributed brain areas are consistently engaged among one another during diverse cognitive tasks \cite{Nak2020, Zen2023}.
Although considered an essential requirement, it is challenging to establish explicit connections among different brain areas.
Specifically, there are critical concerns that need to be addressed, including how the same brain area underlies cognition among varied tasks and how the areas are correlated among divergent human subjects.   
Addressing these questions will reshape our understanding of inner-subject and cross-subject cognitive state paradigms underlying a multi-task mechanism.

Cognition, in essence, refers to the mental action or process of perceiving and reacting, processing and comprehending, focusing and thinking, memorizing and imagining, reasoning and computing, judgment and evaluating, storing and retrieving information, and decision-making and problem-solving \cite{Suh2022}.
Detecting one's cognitive state can help monitor working and memory load and intervene in time, which potentially helps in formulating proper decisions, accomplishing task demands, maintaining mental efficacy, or diagnosing relevant diseases. 
Since cognition generally rely on intelligence, experience, and sensation to betterment one's change in behaviours, its evaluation contrives the diminution brought by in-confidence, information overload, or multitasking.
Insofar, the evaluation of cognitive state is way much more significant for high-load, high-pressure, and high-intensity personnel, such as pilots \cite{Wue2022}.

Flight professionalism-related positions, such as pilots or unmanned aerial vehicle operators, require much more cognitive ability than that of other jobs.
Maintaining a promising cognitive state during tasks is a fundamental yet challenging quality of pilots since it is also indispensable for spatial orientation and flying efficacy.
In the aviation psychology domain, spatial orientation disorders that occur during flight missions (i.e. pilots have confusions and illusions about perceiving their surroundings and positions) are defined as psychological disorders, which are generally caused by decreased cognition and can exacerbate the potentials of accidents \cite{Mos2010}.
More substantially, the concept of ``suicide flight crash'' began to attract public attention in the Germanwings Flight $9525$, and such cases are not unusual.
For example, there are LAM Mozambique Airlines Flight $470$, Japan Airlines DC-$8$, and Royal Air Maroc Flight $630$, to name a few, which are all defined as ``intentional'' or ``deliberately caused by pilot'' \cite{Ken2016}.
Additionally, there are speculations about the causes of the crashes of the Egypt-Air Flight $990$, Malaysia Airlines Flight $370$, and China Eastern Airlines Flight $5735$, but there is still no definitive conclusion to be drawn.
As research highlights that cognition is extensively associated with mental health \cite{Mar2022, Hal2024, Smo2024}, evaluating the cognition state of pilots during flight task is not only crucial for their professional competence, but also critical for avoiding potential disastrous instances.

\begin{figure*}[!t]
\centering
\includegraphics[width=1.8\columnwidth]{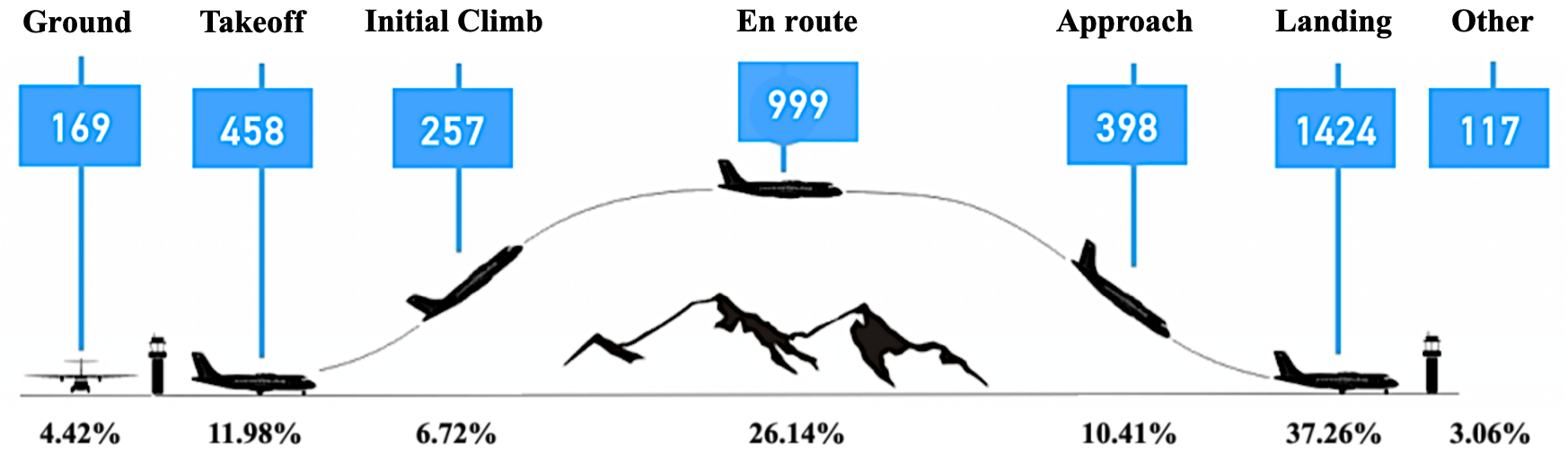}
\caption{$2017$-$2023$ aircraft accidental rates and corresponding occurrence stages (Adapted from \cite{ASN2023}).}
\label{fig1}
\end{figure*}

According to the Aviation Safety Network, there were a total of $3,822$ aviation accidents worldwide from $2017$ to $2023$ (Fig. \ref{fig1}) \cite{ASN2023}, and $70\%$ of those accidents were caused by operational behaviours, and such human errors are closely related to cognitive states \cite{Ngu2019}. 
Yet, pilots are of different ages, genders, and experiences, as well as varied degrees of endurance of fatigue, distraction, stress, workload, task, and information, leading to the distinguished cognition profiles.
Therefore, this study aims to monitor the brain functional connectivity of cognition under multiple tasks for cross-subjects through the use of electroencephalography (EEG) signals since they are objective, non-invasive, cost-efficient, portable, and have high temporal resolution \cite{Suh2022, Kha2023}.
By achieving the objective, this study is the first of its kind which sought to help interpret cognitive connections underlying brain regions, unravel invariant relationships among subjects, capture variability of cognition under the multi-task condition, and evaluate one's cognitive state in real-time, potentially improve pilots' sensitivity to emergencies, accomplish proactive decision-making management, and eliminate safety hazards.
Key contributions of this study are as follows:
\begin{itemize}
    \item This study adopts EEG signals to map brain functional connectivity, analyse its hidden associations underlying cognition, and propose a multi-head attention-based EEGNet for multi-class state evaluation in real-time.
    
    \item Three cognition-related tasks with three different degrees of difficulty are designed to capture subtle changes in brain functional connectivity.

    \item The positively and negatively-related brain functional associations are interpreted from both inner-subjects and cross-subject perspectives to highlight the mutual activated brain regions for uniformity of cognition.

    \item A gender-based functional connectivity analysis is achieved to differentiate the contributing brain regions of cognition, bringing an in-depth and generalised understanding of cognitive state gender-wise. 
    
    \item This study targets the aviation domain, while the proposed methods and findings can be further extended and beneficial to other fields, such as clinics and industry.
\end{itemize}

%contributions can be more technical, maybe include the specific terms for eeg processing, functional connectivity mapping?

\section{Related Works}
\label{sec2}
The evaluation of cognitive state encompasses three approaches: subjective questionnaire-based, task performance-based, and physiological measure-based.
Each method has certain advantages and shortcomings, as described in Table \ref{tab1}.
Specifically, the questionnaire-based approach requires participants to fill self-reports during or after tasks, confronting subjective, resource, and timeliness concerns.
The performance-based approach relies on participants' task measures, which might bring side effects like confrontational mentality or results overemphasis.
The physiology-based approach processes participants' neural or skin signals for monitoring, which is inefficient and has limited interpretability.
Insofar, to establish the brain functional connectivity, physiological measures have been adopted more often.

\begin{table}[!t]
\renewcommand{\arraystretch}{1.3}
\caption{Pros and Cons of Different Cognitive State Evaluation Approaches}
\label{tab1}
\centering
\begin{tabular}{ccc}
\hline
{\bf Methods} & {\bf Pros} & {\bf Cons} \\
\hline
\multirow{3}{*}{Questionnaire-based} & Efficient & Subjective \\
& Accessible & Resources Waste \\
& Uniform &  Limited Timeliness \\
\hline
\multirow{2}{*}{Performance-based} & Proactive & Confrontational Mentality \\
& Objective & Performance Overemphasis \\
\hline
\multirow{2}{*}{Physiology-based} & User-centered & Generalisation Issue \\
& Task-oriented & Limited Interpretability \\
& Objective & Inefficient \\
\hline
\end{tabular}
\end{table}

Yi et al. \cite{Yi2021} adopted EEG signals to construct a large-scale dynamic functional connectivity network by proposing a multivariate correlation analysis wavelet coherence–S estimator in attention-related cognitive tasks.
Zhang et al. \cite{Zhang2021} utilised functional magnetic resonance imaging (fMRI) to construct a third-order network connectivity tensor.
They proposed a novel sparse tensor decomposition method with regularization to factorise the tensor into a series of rank-one components in predicting the cognitive behaviours of individuals.
Similarly, Qu et al. \cite{Qu2022} used fMRI to monitor brain functional connectivity for cross-subject analysis, and they applied graph convolutional networks (GCN) to exploit the relationships for cognition.
Qi et al. \cite{Qi2020} used EEG signals to construct functional connectivity to predict mental fatigue-related behavioural impairments.
Additionally, Sareen et al. \cite{Sar2020} once studied seven intellectual developmental disorder subjects to understand the associations among brain regions during the resting state and listening to music state.
Sen \cite{Sen2021} employed $475$ subjects' fMRI to predict their biological gender and intelligence.
There are many other studies constructing brain functional connectivity for different purposes, such as mapping cognitive simulation, disease detection, and revealing behavioural associations \cite{Yic2022, Xu2024, Tom2024, Hol2024}.

Other than analysing brain functional connectivity, many works are conducted around cognitive state evaluation.
For instance, Zhou et al. \cite{Zho2023} applied the domain adaptation technique to evaluate cross-subject cognitive workload through a maximum mean discrepancy loss, reaching an average accuracy of $74.86\%$.
Chakladar et al. \cite{Das2022} integrated the filter bank common spatial pattern (FBCSP) and long short-term memory (LSTM) to classify cognitive state level using EEG signals, an accuracy rate of $87.04\%$ is proposed under the arithmetic task.
Wang et al. \cite{Wang2022} adopted functional connectivity to monitor and distinguish between cognitive task activation and resting state, and a solid connection was established.
Many more similar studies were derived by adopting physiological measures to evaluate cognition-related states, such as fatigue evaluation \cite{Ma2024}, mental state evaluation \cite{Cui2022}, and emotion recognition \cite{Ren2022}, to name a few.
A brief summary of relevant studies can be found in Table \ref{tab2}.

\begin{table*}[!t]
\renewcommand{\arraystretch}{1.3}
\caption{Cognition-Related Evaluation Literature Sota Comparison}
\label{tab2}
\centering
\begin{tabular}{cccccccc}
\hline
{} & {\bf Data Type} & {\bf No. Instances} & {\bf Pre-processing} & {\bf Functional Connectivity} & {\bf Multi-task} & {\bf Classes} & {\bf Performance} \\
\hline
Zhou et al. \cite{Zho2023} & EEG & $15$ subjects & \ding{51} & \ding{55} & \ding{55} & Binary & $74.07\%$ \\
Li et al. \cite{Lir2022} & EEG & $27$ subjects & \ding{51} & \ding{55} & \ding{55} & Binary & $74.32\%$ \\
Chakladar et al. \cite{Das2022} & EEG & $30$ subjects & \ding{51} & \ding{51} & \ding{55} & Binary & $87.04\%$ \\
Khan et al. \cite{Kha2023} & EEG & $46$ subjects & \ding{51} & \ding{51} & \ding{55} & Binary & $89.98\%$ \\
Ma et al. \cite{Ma2024} & EEG & $15$ subjects & \ding{51} & \ding{55} & \ding{55} & Binary & $92.20\%$ \\
Li et al. \cite{Lij2020} & EEG & $15$ subjects & \ding{51} & \ding{55} & \ding{55} & Multi-class & $88.92\%$ \\
% Cui et al. \cite{Cui2022} & EEG & $19$ subjects & \ding{51} & \ding{55} & \ding{55} & Multi-class & $95.43\%$ \\
{\bf Current Study} & EEG & $30$ subjects & \ding{51} & \ding{51} & \ding{51} & Multi-class & ${\bf95.83\%}$\\
\hline
\end{tabular}
\end{table*}

Despite the existence of literature on cognitive state evaluation, there are limitations draw upon, being: $a$) most of the studies adopt physiological measures to accomplish the goal yet ignoring its shortcomings, $b$) many of the studies ignore the establishment of functional connectivity for inner-subject and cross-subject, $c$) most of the studies are conducted under a simple cognitive task with unitary degree of difficulty, and $d$) most studies tend to work with high and low levels of cognitive state for binary classification while neglecting the transition stage.
Moreover, those methods all confront the challenges of adapting to a new subject.
Accordingly, this study bridges the literature gap and combines the questionnaire-based, performance-based, and physiology-based measures to ensure a fairer evaluation.
Additionally, functional connectivity has been established for inner- and cross-subject analysis, which can provide a better mapping between brain regions and cognition with three distinctive tasks deployed accompanying three degrees of difficulties each.
Upon that, a comprehensive three-level cognitive state has been involved to map the associations subject-wise.

\section{Methodology}
\label{sec3}
With the objective of comprehending cross-subject brain functional connectivity for multi-task multi-class cognitive state evaluation, this section interprets the proposed flight platform, cognitive tasks, and the methodological design.

\subsection{Flight Simulation Platform}
In order to evaluate the cognition of pilots, this study develops a virtual reality (VR)-based flight simulation platform and the cognitive tasks using a $3$D vision and the QT software.
Specifically, the platform consists of a typical routine of a pilot during a flight mission, including route selection, takeoff, en route, approach, and landing.

\begin{figure*}[!t]
\centering
\includegraphics[width=2.0\columnwidth]{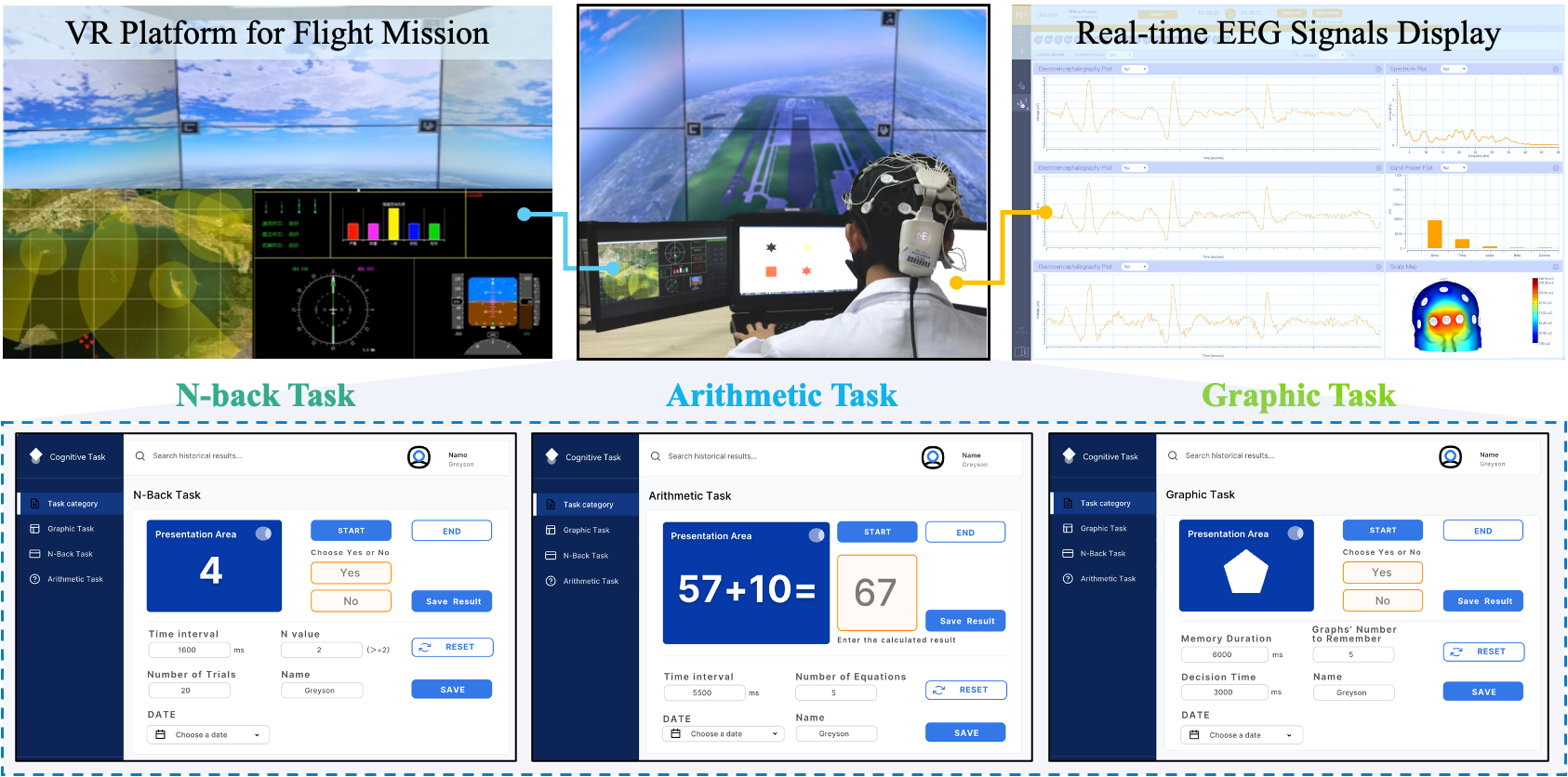}
\caption{Flight simulation platform and the multi-screen cognitive task interface.}
\label{fig2}
\end{figure*}

The functionality of the platform is to offer pilots parallel sub-task operations during the flight mission to induce them to exhibit different levels of cognition. 
In doing so, we use a multi-screen device to display the flight mission scenario, the cognitive tasks, and the pilots' real-time EEG signals (Fig. \ref{fig2}).

\subsection{Cognitive Task Setup}
Cognition is associated with multiple aspects, like memorising, calculating, and reacting \cite{Suh2022, Sen2021}.
In order to comprehensively evaluate those abilities subject-wise, this study proposes three cognitive tasks corresponding to those functionalities displaying in the center of the platform.
Specifically, the tasks are named: $1$) N-back task, $2$) Arithmetic task, and $3$) Graphic Task.
Each corresponds to the three functionalities and is interwoven with each other to evaluate one's information processing, problem-solving, and decision-making capacities.

\subsubsection{N-back task}
The N-back task has been deployed relatively often and has been firmly proven to evaluate one's working memory, sustained attention, and cognitive agility \cite{Qu2022, Cai2024}.
Specifically, the participant will be displayed with a line of numbers in a random sequence one after another, and he/she is required to compare the current stimulus (i.e., number) with the $N$th stimulus that appeared before it.
For example, when $N = 2$, the participant is required to compare the current number with the target number (which is located two stamps before the current position).
If the current and the target stimulus are the same, he/she clicks ``Yes'' on the display, otherwise clicks ``No''.
The participant is required to finish $18$ clicks in one round.
The degrees of complexity and difficulty grow exponentially when we alter the number of $N$ or the reaction time for clicks.
In this case, we fix the number to $N=2$ and shorten the reaction time to $1,600ms$, $1,400ms$, and $1,200ms$ gradually for three difficulty levels to evaluate the participants' working memory in three rounds.

\subsubsection{Arithmetic task}
The arithmetic task is generally applied to test one's reasoning and computing ability and has been adopted in many studies for cognitive state evaluation \cite{Liu2019, Zhang2021, Suh2022}.
Here, we provide addition with two digital numbers for the arithmetic task.
The participant is given the question and required to type in the results in the display and hit the ``Return'' button on the keyboard, or simply leave it until running out of time.
There are $5$, $7$, and $9$ questions in each round when setting the answer time to $5,500ms$, $4,500ms$, and $3,500ms$, which ensures the three rounds with three degrees of difficulty to evaluate one's computing ability.

\subsubsection{Graphic Task}
To evaluate one's attention, visual memory, short-term memory, and reaction capabilities, this study designs a graphic-based task.
Specifically, we involve $5$ shapes (i.e., rectangular, triangle, squares, circle, pentagon, and hexagonal star) of icons with $4$ colours (i.e., blue, red, black, and yellow), which have $20$ types of combinations.
They will be randomly combined and displayed in front of the participant then disappear in a few seconds.
Participants will be required to memorise the appeared graphics and determine whether the newly appeared graphic is in the lastly appeared batch by clicking ``Yes'' or ``No'' buttons on the interface.
We set three degrees of difficulty by displaying $4$, $5$, and $6$ graphics for each round containing $10$ clicks, with a memory duration of $6$ seconds and a judgment interval of $3$ seconds. 
Notably, we ensure the operation time of each task is relatively identical to maintain the fairness.
Detailed information and interface displays can be found in Table \ref{tab3} and Fig. \ref{fig2}.

\begin{table}[!t]
\renewcommand{\arraystretch}{1.3}
\caption{Cognitive Tasks Setup}
\label{tab3}
\centering
\begin{tabular}{ccc}
\hline
{\bf Task} & {\bf Questions} & {\bf Degrees of Difficulty} \\
\hline
N-back: $N=2$ & $18$ & Time = $1600, 1400, 1200ms$ \\
Arithmetic: Add digits & $5, 7, 9$ & Time = $5500, 4500, 3500ms$ \\
Graphic: Icon compare & $10$ & Memorising numbers = $4, 5, 6$\\
\hline
\end{tabular}
\end{table}

\subsection{Dataset Acquisition}
With the ethics obtained from the Medical and Experimental Animal Ethics Committee of Northwestern Polytechnical University, this study acquired $30$ subjects aged between $18$ to $29$ ($\pm$ $2.11$) to conduct the proposed cognitive tasks.
Informed written consents were attained from all participants.
The selection of participants follows a standardised and rigorous protocol that they have to meet the following requirements:
\begin{enumerate}
    % \item All participants are right-handed.
    \item All participants are with normal hearing.
    \item All participants have normal or correct-to-normal vision.
    \item All participants have adequate sleep before experiments.
    \item All participants are asked to avoid strenuous exercise before experiments.
    \item All participants are in good health, with no history of mental or intelligent illness.
\end{enumerate}

\begin{figure}[!t]
\centering
\includegraphics[width=1.0\columnwidth]{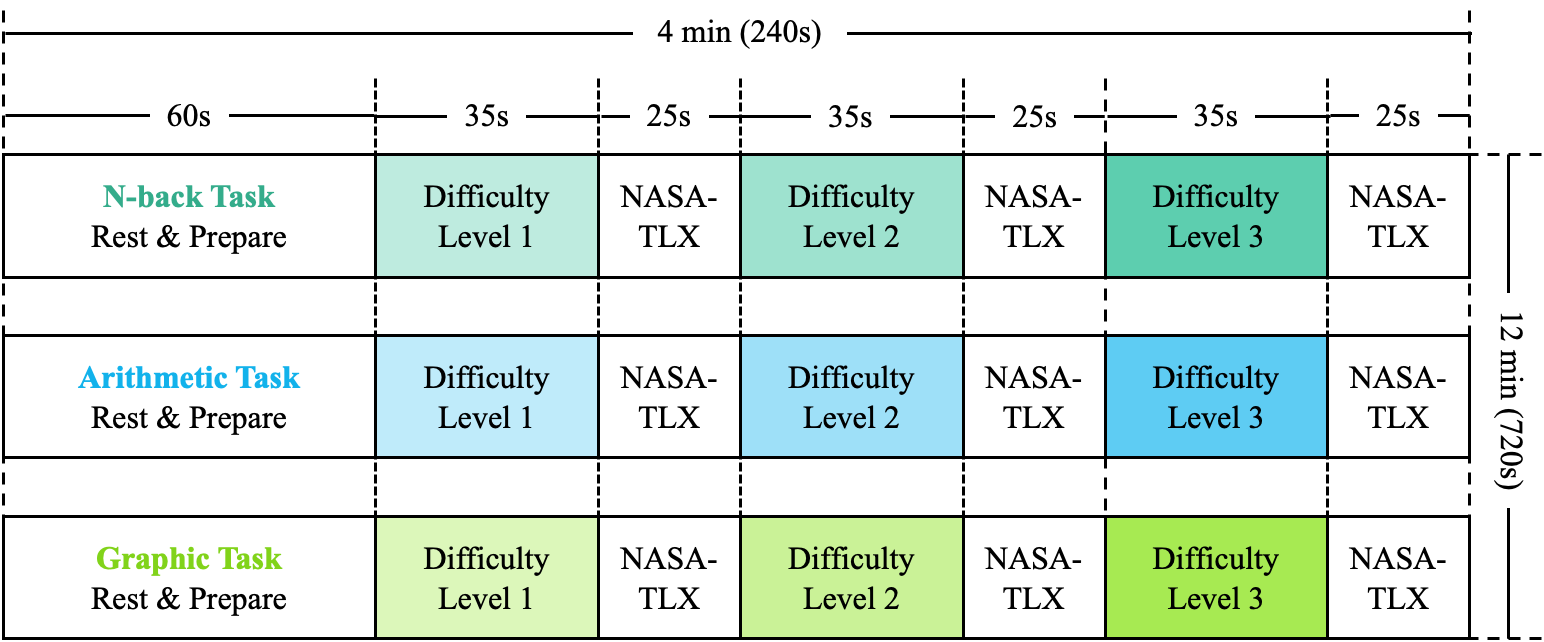}
\caption{Cognitive task experiment setup.}
\label{fig3}
\end{figure}

The detailed participants' demographics and their task scores can be found in Table \ref{tab4}.
Moreover, the overall experiment for one participant is around $12$ minutes, and the detailed experimental setup is illustrated in Fig. \ref{fig3}.

\begin{table}[!t]
\renewcommand{\arraystretch}{1.3}
\caption{Demographics of Participants and Task Performance}
\label{tab4}
\centering
\begin{tabular}{cccc}
\hline
{\bf Measures} & - & {\bf Male (N=$21$)} & {\bf Female (N=$9$)} \\
\hline
Age - Mean\/SD & - & $23.29\pm2.15$ & $23.00\pm1.00$ \\
\multirow{3}{*}{Daytime} & Morning & $28.57\%$ & $22.22\%$ \\
& Afternoon & $52.38\%$ & $22.22\%$ \\
& Evening & $19.05\%$ & $55.56\%$ \\
N-back Task & - & $78.40\%\pm15.20$& $79.84\%\pm12.91$ \\
Arithmetic Task & - & $81.15\%\pm22.80$ & $78.69\%\pm16.31$  \\
Graphic Task & - & $77.94\%\pm15.57$ & $73.33\%\pm22.53$  \\
\hline
\end{tabular}
\end{table}

\subsection{Data Pre-processing}
Since the questionnaire-based, performance-based, and physiological measures-based cognitive state evaluation approaches are complementary with each other, this study combines them for the labelling purpose to ensure a fair evaluation.
Specifically, we use the NASA Task Load Index \cite{Nasa2020} to acquire the participants' self-reports after each task round.
Their task performance and EEG signals are recorded in real-time during experiments.

\begin{figure}[!t]
\centering
\includegraphics[width=1.0\columnwidth]{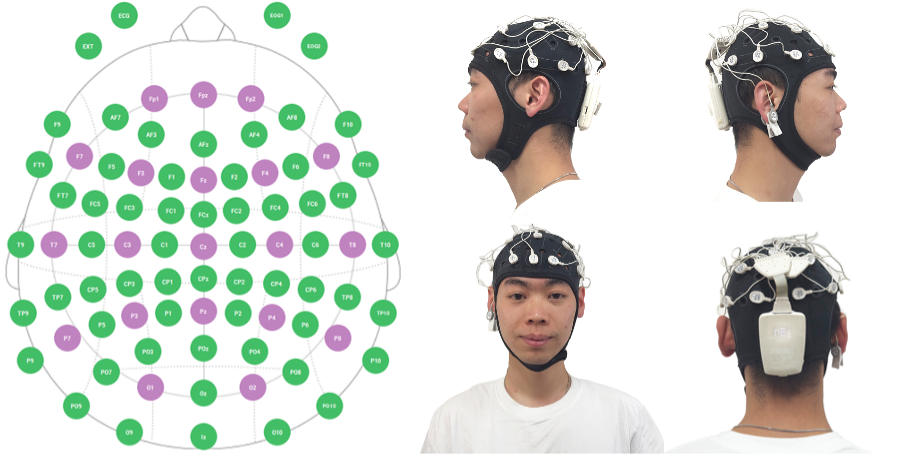}
\caption{Utilised $20$-lead EEG electrode system.}
\label{fig4}
\end{figure}

In this study, we utilise the $20$-lead dry electrodes for EEG signal acquisition with an NE EEG cap with signals acquired and stored through the NIC$2$ software (refer to Fig. \ref{fig4} where purple leads are the adopted ones).
The utilised electrodes are: Fp$1$, Fpz, Fp$2$, F$7$, F$3$, Fz, F$4$, F$8$, T$7$, C$3$, Cz, C$4$, T$8$, P$7$, P$3$, Pz, P$4$, P$8$, O$1$, and O$2$.
Each electrode placement site can be read from: prefrontal (Fp), frontal (F), temporal (T), parietal (P), occipital (O), and central (C). 
Before conducting the cognitive tasks, participants are asked to sit immobile for device setup, including electrode installation, unilateral earlobe electrode referencing, electrode configuration, and signal quality inspection.
After the initial setup, the participant can then start the tasks and their EEG signals are recorded and processed subsequently.

\begin{figure*}[!t]
\centering
\includegraphics[width=2.0\columnwidth]{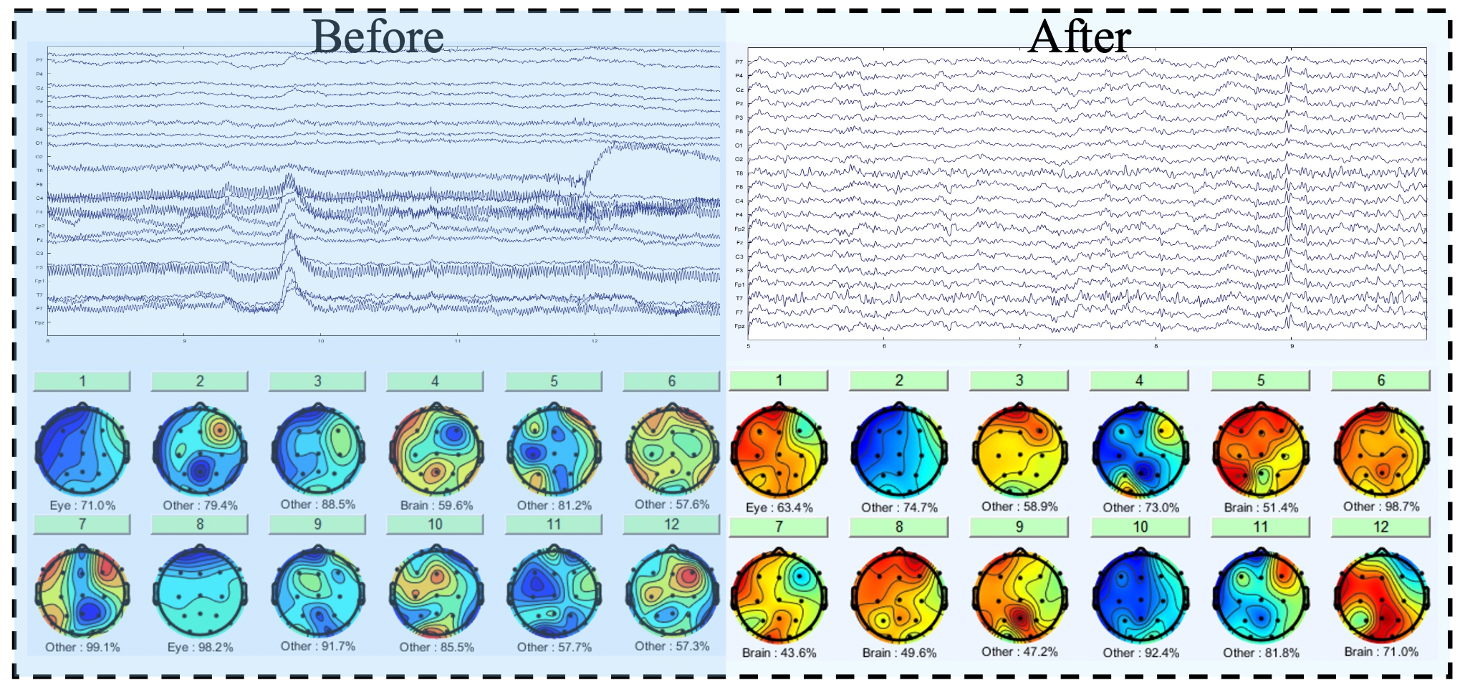}
\caption{EEG signals and brain electrical activity maps before and after the pre-processing steps.}
\label{fig5}
\end{figure*}

We use the EEGLab toolbox in MATLAB to pre-process the raw EEG signals, which includes electrode positioning, previewing, interpolation, filtering, baseline correction, and independent component analysis (ICA) for removing artefacts \cite{Wen2022}.
Fig. \ref{fig5} demonstrates the raw and the pre-processed EGG signals.
We set the sampling frequency of $0$ to $500$Hz.
During the signal acquisition, there will be corrupted data caused by poor electrode contact, which leads to obvious waveform drifts or artefacts, thus, they are filtered and interpolated to get it recovered.
We use a band-pass filter of $0.1$ to $50$Hz and a $46$ to $50$Hz band-stop filter to process the signals.
Since there will be irrelevant electrical signals caused through the electromyographic effects of blinking or eye movements, they are corrected through ICA.
Then, the processed signals are separated into five different frequency bands, including $\delta$ ($0.1$-$4$Hz), $\theta$ ($4$-$8$Hz), $\alpha$ ($8$-$13$Hz), $\beta$ ($13$-$30$Hz), and $\gamma$ ($30$-$50$Hz).
 
Before mapping the activity flow among the $20$ channels, the EEG signals are normalised into a standardised and consistent scale through Equation \ref{eq1}, where $x_i$ is the $i$th signal in electrode $x$.
Pre-processing details are interpreted in Algorithm \ref{alg1}.

\begin{equation}\label{eq1}
    x_i^{'} = \frac{x_i-min(x)}{max(x)-min(x)}
\end{equation}

\begin{algorithm}[t]
\label{alg1}
\caption{EEG Signals Pre-processing Procedures}
{\bf Input:} Raw EEG signals $X_i^t$ \\
{\bf Output:} Pre-processed EEG signals \\
Initial: Set $i=1$, $K=20$, $t=0$; $i$ is the $i$th electrode, $K$ is the number of electrodes, $t$ is the signal time duration \\
    \While{$i$ in $K$ electrodes}{
        Preview $X_i^t$ \\
        \If{$X_i$ is corrupted}{
            \eIf{$X_i^0$ to $X_i^t$ is corrupted}{Interpolate the corrupted signals}{Interpolate the corrupted segments} 
        }
        Filter $X_i$ with a band-pass filter with a range of [$0.1Hz$, $80Hz$] \\
        Filter $X_i$ with a notch filter with a range of [$49Hz$, $51Hz$] \\
        Apply Baseline Correction with a range of [$3000ms$, $5000ms$] \\
        Apply Independent Component Analysis \\
        $i=i+1$ \\
    }
\end{algorithm}

\subsection{Functional Connectivity Construction}
Functional connectivity can depict synchronous associations of information flow between two EEG signal channels \cite{Suh2022, Xu2024}.
In this study, we use the Pearson's correlation coefficient (PCC) to quantify the associations among channels in a time series sequence.
PCC is calculated through Equation \ref{eq2}.
It generates an association degree ranging from $-1$ to $1$, indicating whether the association is negatively or positively related, respectively.

\begin{equation}\label{eq2}
    PCC = \frac{\sum_{i=1}^{n}(x_i-\bar{x})(y_i-\bar{y})}{\sqrt{\sum_{i=1}^{n}(x_i-\bar{x})^2}\sqrt{\sum_{i=1}^{n}(y_i-\bar{y})^2}}
\end{equation}

Specifically, $n$ is the number of time-series segments, $x_i$ is the $i$th signal in electrode $x$, $\bar{x}$ is the averaged signals in $x$, $y_i$ is the $i$th signal in electrode $y$, and $\bar{y}$ is the averaged signals in $y$.
We use the $20$-lead electrode EEG cap to record signals, thus, the adjacency matrix representing functional connectivity correlation is constructed and has a size of $20\times20$ for each subject.
Then, all of the $30$ subjects' functional connectivity correlations are calculated.
By sorting the correlations among channels in the descending order for each subject, we then get a $N \times C \times T$ correlation embedding, where $N$ is the number of participants, $C$ is the number of EEG signal channels, and $T$ is the number of tasks.

In order to have a comprehensive analysis of the cross-subject functional connectivity, we combine the embedding through different levels, including overall combination, task-level combination, and gender-level combination.
Specifically, we add the correlations among channels for all the subjects to get the overall connectivity flow.
For the task-level, we assign weights to the degrees of task difficulty based on the participants' performance using Equation \ref{eq3}, where $i$ is the $i$th correlation between channels, $w_i$ is the weight to be assigned for each difficulty level, $\mathcal{P}_{\bar{T_i}}$ is the average accuracy performance for a specific level given a task, and $n$ is the total number of difficulty levels.
Then the connectivity correlations are combined through the weighted addition.
As for the gender-level, the functional connectivity correlations are combined for males and females separately.

\begin{equation}\label{eq3}
    w_i = \frac{\mathcal{P}_{\bar{T_i}}}{\sum_{i=1}^n{\mathcal{P}_{\bar{T_i}}}}
\end{equation}

We notice that many of the subjects' correlation matrices encounter negative values, through addition and weighting, the highly associated connectivity flows are all with positive values and are further adopted for analysis.

\subsection{Mental State Evaluation}
With the sequential correlations among subjects, a set of highly influential electrodes are identified, and the EEG signals of those selected electrodes are given as input to the classifier for real-time mental state evaluation.
The classification is performed among the multi-class cognitive states, including low, high, and transition stages.

\begin{figure}[!t]
\centering
\includegraphics[width=0.95\columnwidth]{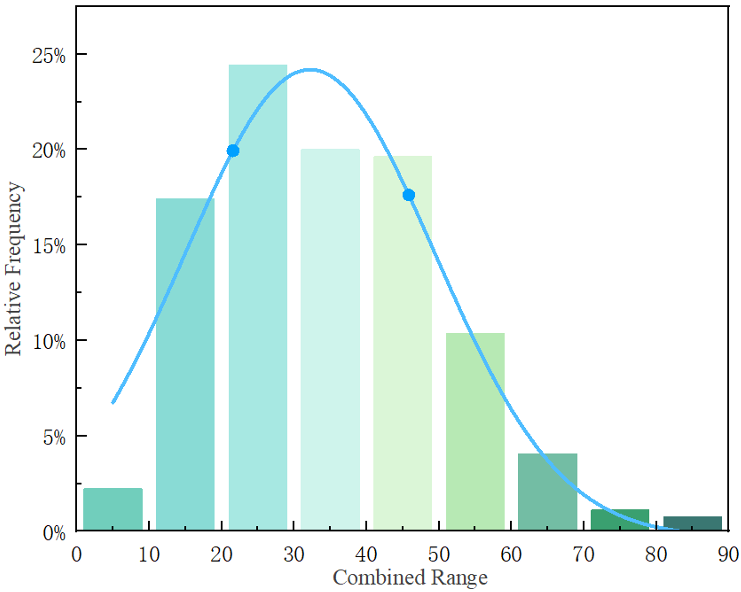}
\caption{Multi-class cognitive state sample distribution.}
\label{fig6}
\end{figure}

This study utilises the NASA-TLX self-reports and the task performance to label the EEG signals for the multi-class cognitive states.
Specifically, the Quartile method is applied to identify the range for each class, and then the report result and the performance are averaged to determine the class for each subject by fitting into the Quartile ranges.
The distribution of the classes can be viewed in Fig. \ref{fig6}.
Additionally, this study provides partial dataset and full code access in the GitHub link \url{https://github.com/d-lab438/Multi-channel-eegnet.git}.

\begin{figure}
    \centering
    \includegraphics[width=1\columnwidth]{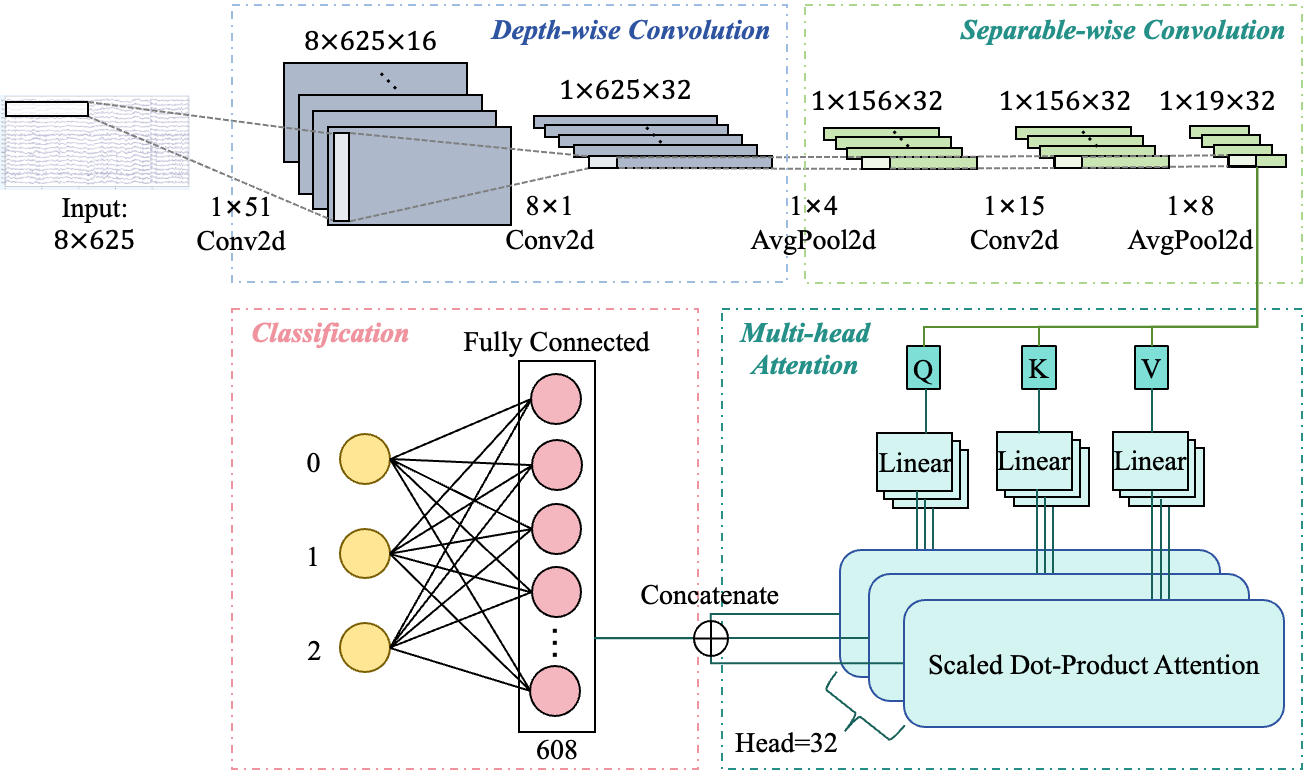}
    \caption{The proposed MHA-EEGNet architecture.}
    \label{fig7}
\end{figure}

We adopt the most commonly used classifiers for cognitive state evaluation, including Random Forest (RF), Multi-layer Perceptron (MLP), Long Short-Term Memory (LSTM), and EEGNet.
Additionally, we propose a Multi-head Attention-based EEGNet (MHA-EEGNet) for generating precise multi-class evaluation results. 
The proposed MHA-EEGNet architecture is depicted in Fig. \ref{fig7}, which extends the conventional EEGNet by integrating multi-head attention module following separable convolutions. 
This addition module enables the network to selectively focus on relevant spatial and temporal features across different channels and time steps of EEG data, facilitating more discriminative feature representation for subsequent classification tasks.
By dynamically learning attention weights, the model enhances its sensitivity while maintaining robustness against noise and irrelevant signals, thereby improving evaluation accuracy.

For all the experiments, the stratified $10$-fold cross-validation (CV) is applied, which allows to maintain the class distribution ratio in both training and testing splits during each iteration.
This can help compensate for the unbalanced class distribution and mitigate the accuracy variance.
We set the batch size to $10$.
Additionally, the categorical cross-entropy (CCE) is applied to generate classification results for the multi-class scenario, and the calculation is made through Equation \ref{eq4}, where $s_p$ denotes the predicted value for a target class, $t_q$ is the inferred score for each class in $c$, while $c$ is the total number of classes.

\begin{equation}\label{eq4}
    CCE = -log({\frac{e^{t_p}}{\sum_{q}^{c} e^{t_q}}})
\end{equation}

Notably, the multi-class classification is conducted subject-independently to validate the identified functional connectivity.
Through the CV results, the robustness and generalisation of the model can be validated.
Insofar, as the signals from the $20$-lead electrodes are also input into the classifiers as baseline results, which are compared to the results of inputting the selected electrodes.
The performance is evaluated through six metrics, and they are calculated through Equations \ref{eq5} to \ref{eq10} ($TP$ stands for ``True Positive'', $TN$ denotes ``True Negative'', $FP$ is ``False Positive'', $FN$ is ``False Negative'', $NPV$ is ``Negative Predictive Value'', and $k$ is the number of folds).

\begin{equation}\label{eq5}
    Accuracy = \frac{1}{k} \sum\nolimits_{i=1}^{k} {\frac{TP_i+TN_i}{TP_i+TN_i+FP_i+FN_i}}
\end{equation}

\begin{equation}\label{eq6}
    Precision = \frac{1}{k} \sum\nolimits_{i=1}^{k} {\frac{TP_i}{TP_i+FP_i}}
\end{equation}

\begin{equation}\label{eq7}
    Recall = \frac{1}{k} \sum\nolimits_{i=1}^{k} {\frac{TP_i}{TP_i+FN_i}}
\end{equation}

\begin{equation}\label{eq8}
    Specificity = \frac{1}{k} \sum\nolimits_{i=1}^{k} {\frac{TN_i}{TN_i+FP_i}}
\end{equation}

\begin{equation}\label{eq9}
    NPV = \frac{1}{k} \sum\nolimits_{i=1}^{k} {\frac{TN_i}{TN_i+FN_i}}
\end{equation}

\begin{equation}\label{eq10}
    F1 = 2 \times {\frac{Precision \times Recall}{Precision + Recall}}
\end{equation}

In addition, all the experiments are conducted under the same computational environment, which is through a $64$-bit Windows $11$ Pro desktop, which has an $13$th Gen Intel Core i$7$-$13650$HX processor with $16$ gigabytes of memory and a Geforce RTX $4060$ GPU.

\section{Results and Discussion}
\label{sec4}
This section elaborates on the brain functional connectivity from three perspectives, including overall, task-level, and gender-level.
Additionally, the multi-class cognitive state classification performance is illustrated to verify the associations.

\subsection{Overall Analysis}
\begin{figure*}[!t]
\centering
\includegraphics[width=2.0\columnwidth]{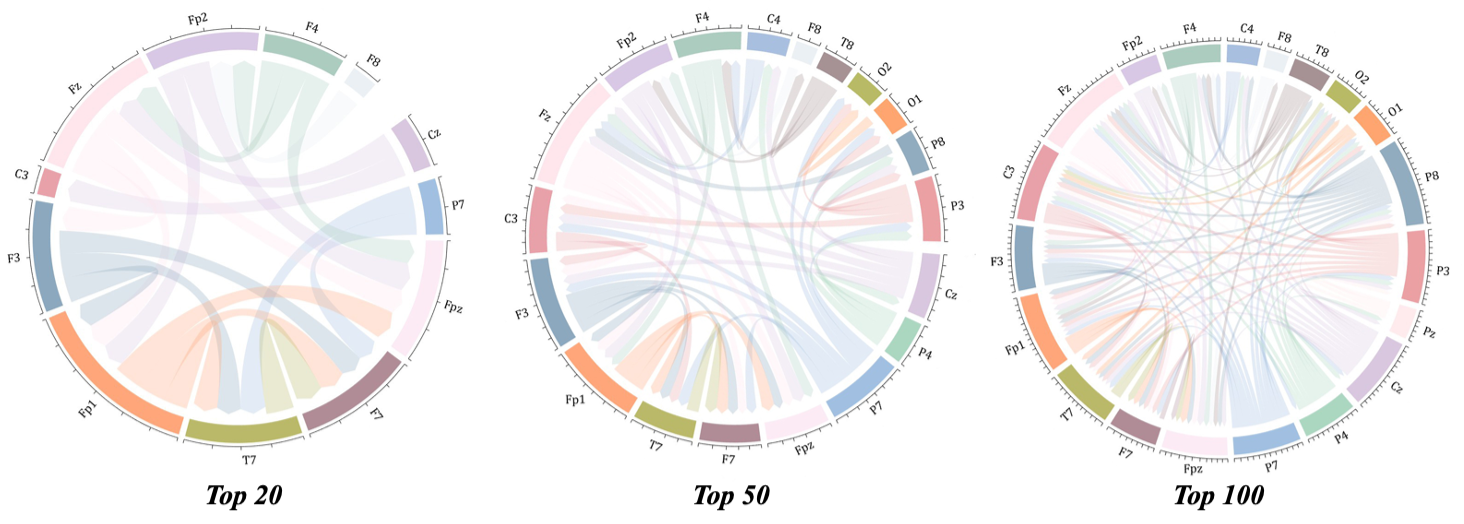}
\caption{Overall subject-independent chord diagram of functional connectivity using the top $20$, $50$, and $100$ correlations.}
\label{fig8}
\end{figure*}

\begin{figure}[!t]
\centering
\includegraphics[width=1.0\columnwidth]{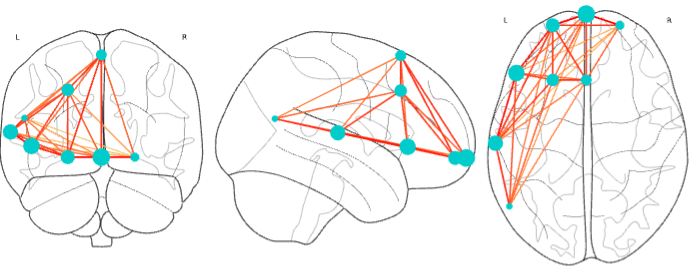}
\caption{Overall subject-independent brain functional connectivity map.}
\label{fig9}
\end{figure}

By performing the pre-defined aviation and cognitive tasks, the participants' functional connectivity is identified and stored in a sequential list.
Fig. \ref{fig8} demonstrates the chord diagram of the subject-independent functional connectivity by adopting the top $20$, $50$, and $100$ associations for all participants.
The chord diagram is applied to illustrate the brain's electrical signal connections among electrodes.
It is evident that $Fz$ is the most significant electrode as it takes the widest segments, showing strong associations with other points like Fp$1$, Fpz, and F$3$.
To better understand the functional connectivity concretely, Fig. \ref{fig9} exhibits the connected electrodes underlying a brain map.
Based on the demonstration, the associations are majorly located on the left hemisphere, and the frontal lobe is the most active region, including electrodes Fp$1$, Fpz, Fp$2$, F$7$, Fz, and F$3$.
Under the lower left side, electrodes T$7$ and P$7$ from the temporal and parietal lobes are responsive.

The left hemisphere generally controls language processing and intellectual functions, which are related to reading, lexical retrieval, speech, logic, reasoning, writing, comprehension, and analytics \cite{Her2020}.
The frontal lobe and prefrontal cortex demonstrate sophisticated interwoven, which indicates their solid roles in decision-making and semantic processing.
Since the three tasks are associated with working memory, arithmetic, and visuospatial cognition, the connected brain regions are in accordance with the findings stated in \cite{Her2020, Mes2023}.
As for the temporal lobe, it has established connections with the frontal and parietal lobes in the left hemisphere, highlighting its characteristic in processing short-term memory.
Additionally, it has been pointed out that the functional interactions between the prefrontal cortex and the medial temporal structures are crucial for encoding and retrieving \cite{Her2020}.
This statement has been verified through our experiments, where the prefrontal cortex, frontal lobe, partial temporal, and parietal lobes are connected, in accordance with the well-established dominance for information retrieving, comprehending, and processing.
However, the temporal and parietal lobes are not so much involved in the defined tasks, this might be due to the responsibilities of the two lobes which are generally involved with rhythm, smell, objects and spatial relationships identifications, interpreting pain and touch, and also spoken language comprehension \cite{Dir2021, Hab2022}. 
In this case, we highlight the cognition-related tasks, therefore, the sensory-related abilities (e.g., auditory, somatosensory) are not evaluated.

\subsection{Task-level Analysis}
\begin{figure*}[!t]
\centering
\includegraphics[width=2.0\columnwidth]{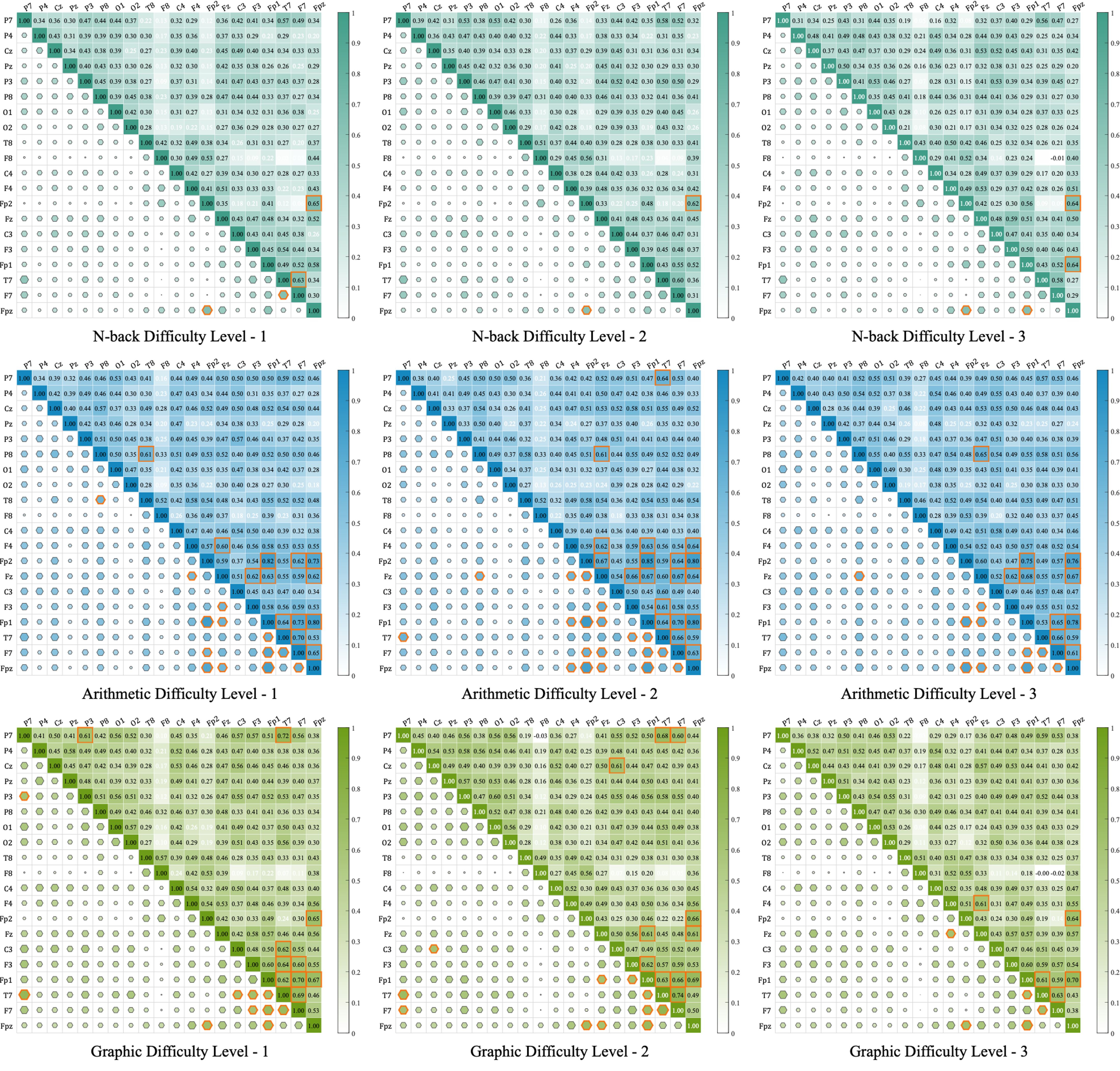}
\caption{Overall subject-independent task level-based brain functional correlations among the $20$ electrodes.}
\label{fig10}
\end{figure*}

In order to capture the subtle changes in brain functional connections in varied degrees of difficulty for the three tasks, here we interpret the correlations through task-level analysis.
Table \ref{tab5} presents the average performance of the three tasks.
Fig. \ref{fig10} displays the correlation matrix for the three tasks embedded with three difficulty levels.

\begin{table}[!t]
\renewcommand{\arraystretch}{1.3}
\caption{Average Performance of The Three Cognitive Tasks}
\label{tab5}
\centering
\begin{tabular}{cccc}
\hline
{\bf Tasks} & {\bf Difficulty} & {\bf Task Performance} & {\bf NASA-TLX}\\
\hline
\multirow{3}{*}{N-back} & Level $1$ & $83.15\%$ & $47.70\%$ \\
& Level $2$ & $80.74\%$ & $56.37\%$ \\
& Level $3$ & $72.59\%$ & $66.87\%$ \\
\hline
\multirow{3}{*}{Arithmetic} & Level $1$ & $87.33\%$ & $43.90\%$ \\
& Level $2$ & $84.29\%$ & $54.30\%$ \\
& Level $3$ & $69.63\%$ & $67.23\%$ \\
\hline
\multirow{3}{*}{Graphic} & Level $1$ & $86.00\%$ & $45.20\%$ \\
& Level $2$ & $75.33\%$ & $55.97\%$ \\
& Level $3$ & $68.33\%$ & $73.42\%$ \\
\hline
\end{tabular}
\end{table}

For the N-back task, the majorly connected brain regions are prefrontal, frontal, and temporal.
It is intriguing to also notice that the frontal and temporal lobes are relatively influential in the first level of difficulty, whereas the prefrontal cortex is significant in all levels of difficulty.
Under such a scenario, the prefrontal cortex exhibits a high impact on the task, suggesting that it might play a pivotal role in working memory.
This assumption is consistent with the findings proposed in \cite{Bre2007} where they found memory suppression may occur in non-psychiatric sample populations under the control of prefrontal regions.
However, the temporal lobe has established connections with the frontal lobe, demonstrating activation in the semantic and speech-related processing, which has been shown in \cite{Hen2021}.
This might be due to the fact that the N-back is the first task in the experiment, and we generally ask the participant to practice for a few rounds.
In this case, they are allowed to ask questions, leading to the apparent activation in the temporal and frontal lobes.
This kind of activation gradually fades as the experiment initiates, therefore, the connections are not evident in the second and third levels of difficulties.

The arithmetic task establishes more solid connections between the prefrontal and frontal lobes. 
In addition, the second degree of difficulty exhibits the most connectivity, suggesting participants are paying more attention and requiring more cognitive resources in this case.
Yet, when they move on to the next difficulty level, the connections start to dribble away.
This might be because the answer time for the third level is rapid, and the participant does not have sufficient time and effort in computing, therefore, many of them might decide to input random answers.
Based on the average performance for task two in Table \ref{tab5}, the assumption is supported, as this is also the task which shows the biggest decrease from level $2$ to level $3$ of around $15\%$ accuracy rate.

The graphic task demonstrates intriguing connections where not only the left hemisphere is involved, but also the partial right hemisphere is responsive.
Besides prefrontal, frontal, parietal, and temporal lobes, the central region is also activated, and the electrodes C$3$, Cz, and F$4$ have been identified as associative during the task.
In the first two levels, the central region is more engaged, whereas the right frontal lobe begins to activate in the third trial.
The right hemisphere controls creativity, spatial ability, artistic, musical skills, imagination, and visual memory, thus, it is also known as the artistic brain \cite{Gee2018, Chen2020}.
In this case, the icons appearing in the interface are increasing in the three iteration, and this might provoke the stimulus of a stabilised visualisation functionality and memory.
By observing the correlation matrix and the average performance of task and NASA-TLX, the participants rate the graphic task as the most pressuring one and indicating it requires extensive cognitive loads.
Evidently, the more brain regions are activated, the more cognitive loads are consumed, and the lower self-estimations are emerged.

\subsection{Gender-level Analysis}
\begin{figure}[!t]
\centering
\includegraphics[width=0.9\columnwidth]{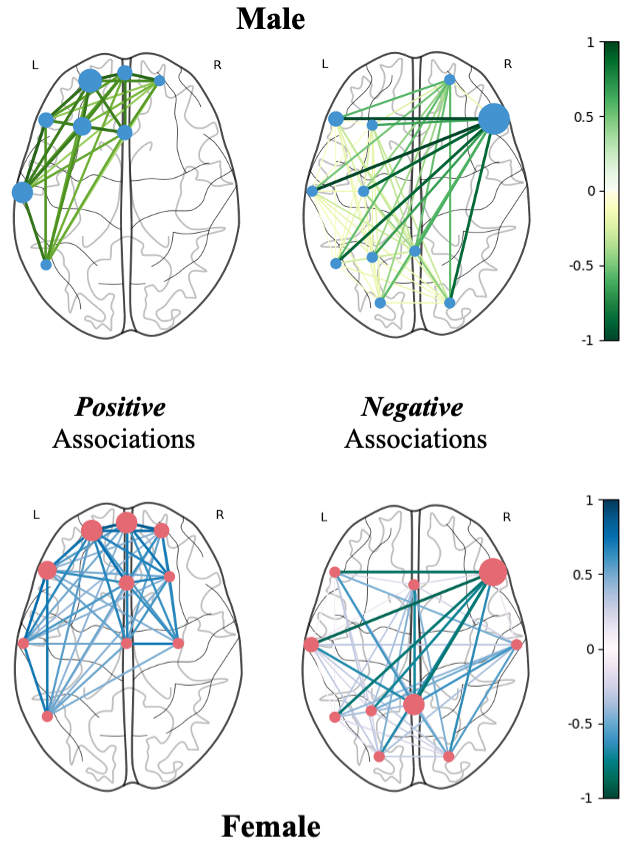}
\caption{Gender-based positively and negatively-related functional connectivity: top presents the positive and negative associations for the male group, and below presents the positive and negative associations for the female group.}
\label{fig11}
\end{figure}

During the inner-subject and cross-subject brain functional connectivity analysis, we notice that many of the trials are embedded with negative values, indicating participants confronting abnormal brain activation during experiments.
Accordingly, here we interpret the brain functional connectivity for males and females through both positively- and negatively-related correlations.
Fig. \ref{fig11} shows the connected regions.

Unsurprisingly, the male group's positively related functional connectivity is consistent with the overall result, which is due to the number of males being much higher than the female group.
When considering the female group separately, the connectivity varies to a noticeable extent.
Besides the Fp$1$, Fpz, Fp$2$, F$7$, Fz, T$7$ and P$7$ electrodes, F$4$, Cz, and T$8$ are also inter-connected.
Differing from the male group, the female group tends to involve more brain region activation in the right hemisphere.
However, this finding is controversial with \cite{Tia2011} where they found the male group generally had a higher local activation in the right hemisphere and the female group had a higher local activation in the left hemisphere.
Nevertheless, their work was conducted under the resting-state scenario, whereas our study was performed under the task-state, thus, the divergence is understandable and explicit, yet intriguing. 
Notably, \cite{Lia2021} indicated that there is a higher leftward asymmetry in the male group than the female group, which aligns with our findings.
The connectivity map also suggests that the female group has more connections and they are generally crisscross across the central gyrus. 
This statement has been supported in \cite{Sen2021} and \cite{Chr2018}, where they stated that the intra-hemispheric connectivity is greater in male brains and inter-hemispheric connectivity is greater in female brains. 
In addition, the wider regions involved in the right hemisphere of females highlight their superiority in object detection and episodic memory, and this finding is in accordance with \cite{Sem2020}.

Considering the negative correlations for males and females, the activation of the left hemispheric regions is more predominant in the male group than in the female group.
For the male group, negative functional connectivity is majorly seen in the left frontal to parietal regions with the middle occipital gyrus involved.
This finding is in consistency with the statements proposed in \cite{Chr2018} where the male group tends to produce higher functional connectivity to the right insula (i.e., located between the frontal and temporal lobes), the left uncus (i.e., located in the temporal lobe), and the left medial orbital prefrontal cortex.
For the female group, negative functional connectivity is relatively in the centralised and lateral regions, which can be seen in nearly all the lobes, and mostly activated in the post-central gyrus and the mid-cingulate cortex.

\begin{figure*}[!t]
\centering
\includegraphics[width=2.0\columnwidth]{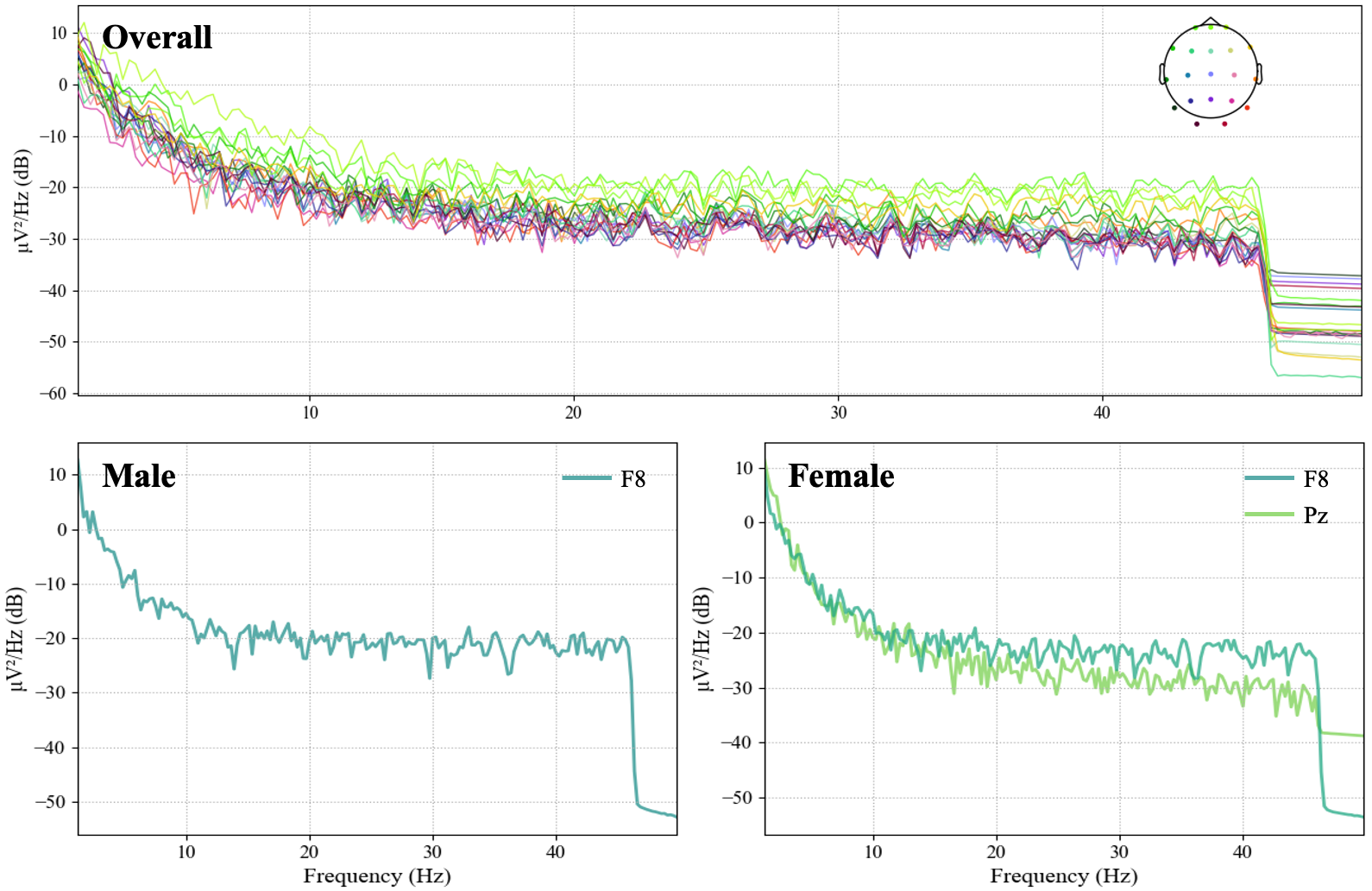}
\caption{Power spectral density demonstration: top shows the density change for all subjects, and below demonstrates male and females' band frequencies.}
\label{fig12}
\end{figure*}

In addition, we decompose and calculate the power spectral density of the mostly activated positive and negative associations for both the male group and the female group.
Fig. \ref{fig12} displays the overall five bands' activities change.
Specifically, the mostly engaged electrodes for males are F$8$, while F$8$ and Pz are more engaged for females.
Based on the analysis, the $\delta$ band has the highest activity during the conduction of the experiment, which is in accordance with \cite{Mai2023}.

\subsection{Cognitive State Evaluation}
\begin{figure}
    \centering
    \includegraphics[width=1.0\linewidth]{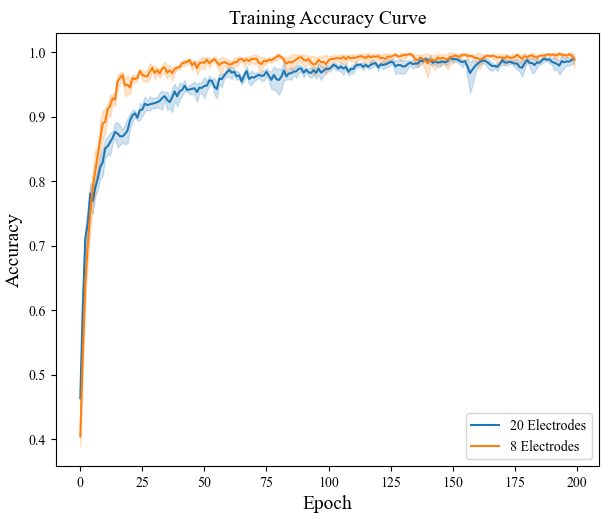}
    \caption{MHA-EEGNet training curves ($200$ epochs).}
    \label{fig13}
\end{figure}

In order to reach a relatively generalised cognitive state evaluation, we have identified $8$ electrodes (i.e., Fp$1$, Fpz, Fp$2$, F$7$, F$3$, Fz, T$7$, and P$7$) which are significant and can capture subtle cross-subject association changes. 
Therefore, we utilise the real-time EEG signals from the $20$ electrodes as input to the classifiers as comparative study.
The generated performance is then compared to the performance when only inputting the signals from the selected $8$ electrodes.
Table \ref{tab6} and Fig. \ref{fig13} display the detailed comparative performance.

\begin{table*}[!t]
\renewcommand{\arraystretch}{1.3}
\caption{Performance Comparison of Inputting the $20$ and $8$ Electrodes' Signals (six metrics from accuracy to F$1$)}
\label{tab6}
\centering
\begin{tabular}{llcccccc}
\hline
{\bf Input Signals} & {\bf Model} & {\bf Accuracy} & {\bf Precision} & {\bf Recall} & {\bf Specificity} & {\bf NPV} & {\bf F$1$} \\
\hline
\multirow{5}{*}{$20$-electrodes}& Random Forest & $0.7583$ & $0.7812$& $0.7347$ & $0.8722$ & $0.8859$ & $0.7447$ \\
& Multi-layer Perceptron & $0.7833$& $0.7638$ & $0.7697$& $0.8943$& $0.8905$& $0.7645$\\
& Long Short-Term Memory & $0.7750$ & $0.7615$ & $0.7522$ & $0.8835$ & $0.8861$ & $0.7559$ \\
& EEGNet & $0.7583$ & $0.7541$ & $0.7297$ & $0.8749$& $0.8840$& $0.7334$\\
& Proposed MHA-EEGNet & ${\bf0.8583}$& ${\bf0.8731}$& ${\bf0.8220}$& ${\bf0.9237}$& ${\bf0.9336}$&  ${\bf0.8366}$\\
\hline
\multirow{5}{*}{$8$-electrodes}& Random Forest & $0.8667$ & $0.8755$ & $0.8556$ & $0.9304$ & $0.9349$ & $0.8619$ \\
& Multi-layer Perceptron & $0.8667$ & $0.8563$ & $0.8530$ & $0.9321$ & $0.9339$ & $0.8537$ \\
& Long Short-Term Memory & $0.9167$ & $0.9141$ & $0.9185$ & $0.9576$ & $0.9578$ & $0.9148$ \\
& EEGNet & $0.8833$ & $0.8748$ & $0.8496$ & $0.9395$ & $0.9451$ & $0.8584$\\
& Proposed MHA-EEGNet & ${\bf0.9583}$ & ${\bf0.9671}$ & ${\bf0.9415}$ & ${\bf0.9770}$ & ${\bf0.9819}$ & ${\bf0.9516}$ \\
\hline
\end{tabular}
\end{table*}

\begin{figure}[!t]
\centering
\includegraphics[width=1.0\columnwidth]{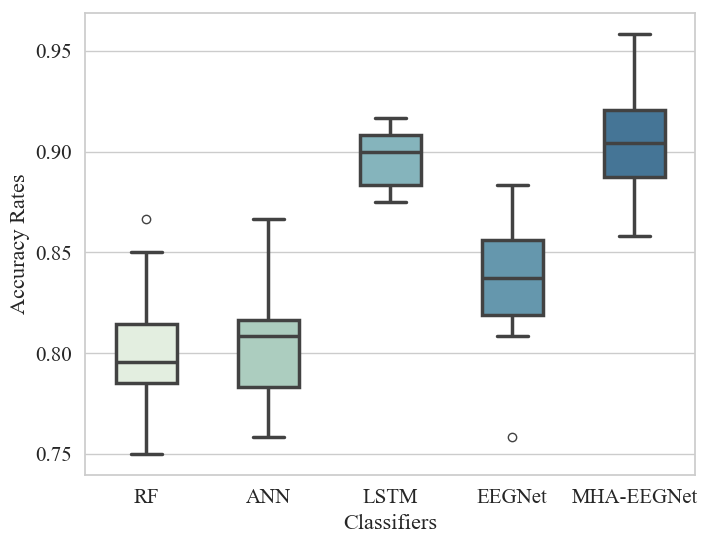}
\caption{Boxplot for the five classifiers' performance with the selected $8$ electrodes signals.}
\label{fig14}
\end{figure}

Based on the comparative performance, it is evident that by adopting the signals from the selected $8$ electrodes, the multi-class classification performance increases.
Specifically, all the classifier demonstrate comparative and even superior accuracy rates.
Considering the precision, recall, specificity, and NPV scores of those classifiers for both input, we can learn that the model generats more stable results.
The potential assumption is that the overall $20$ electrodes might contain non-contributing associations (i.e., noises) which will affect the ultimate evaluation performance.
Therefore, by removing the redundant and irrelevant information, the classification performance is enhanced of around $8.34\%$ to $22.50\%$.
We suggest using the selected $8$ left-side electrodes for cross-subject cognitive state evaluation in future relevant research.
More importantly, Fig. \ref{fig14} exhibits the performance comparison among the five classifiers.
The proposed MHA-EEGNet generates better results in both $20$ and $8$ electrodes' datasets, indicating its robustness in multi-class cognitive state evaluation.  

\section{Conclusion}
\label{sec5}

This study targets the aviation domain for pilot cognitive state evaluation by incorporating the analysis of inner-subject and cross-subject brain functional connectivity through EEG signals.
We design three cognitive tasks each with three levels of difficulty, embedded within the VR-based flight platform.
In our work, the influences of task differences, difficulty level differences, and gender differences on brain functional connectivity are all examined to reach better interpretability and cross-subject generalisation of brain connections.
Key findings indicate that the left hemisphere is more actively engaged in the task-state cognition, different tasks induce varied brain region activation mainly around the prefrontal and frontal lobes, and females tend to exhibit wider brain region connections for the inter-hemisphere than males.
Additionally, we find that the $\delta$ band is the most significant frequency band in both male and female groups underlying negative cognition correlations.
Based on the findings, this study extracts $8$ key electrodes for real-time multi-class cognitive state evaluation.
We further propose a MHA-EEGNet model for precise cognitive state evaluation.
A $10$-fold averaged accuracy of $95.83\%$ is achieved, reaching the state-of-the-art performance.

Although this study is the first of its kind which monitors brain functional connectivity using multi-task, multi-degree of difficulty, and multi-class context for cross-subject cognition evaluation, certain limitations lie underneath.
One of the main concern is related to the negative correlations, which are not analysed task-wise, and this might bring assumptions on whether different types of tasks will provoke negative associations more easily.
Additionally, since all of our participants are right-handed, whether the left-handed subjects will experience similar or opposite findings also needs to be examined.
Moreover, the experimental time during the day has been recorded for each participant, yet its association with cognition is not analysed and interpreted in this study due to scope restrictions.
Therefore, in future work, we will expand our sample populations and implement more comprehensive analysis to understand cognition, emotion, and mental states.
Together, we envision the findings derived from this study provide foresight into cross-subject cognitive functional connectivity mapping and comprehension, which will be beneficial for aviation safety and brain science.

\section*{Declaration of competing interest}
The authors declare that they have no known competing financial interests or personal relationships that could have appeared to influence the work reported in this paper.

\section*{Acknowledgment}
The authors would like to sincerely thank the participants.
This study is supported by the Postdoctoral Fellowship Program of China Postdoctoral Science Foundation under Grant Number GZB$20240985$ and the National Natural Science Foundation of China (No. $61305133$, $52372398$).

\bibliographystyle{ieeetr}

% \balance 
%\vfill

\end{document}